\documentclass{Interspeech2024}
\usepackage{multirow}
\usepackage{bibspacing}

\interspeechcameraready

\title{Codecfake: An Initial Dataset for Detecting LLM-based Deepfake Audio}

\small \name[affiliation={1,2,\dagger}]{Yi}{Lu}
\small \name[affiliation={1,3,\dagger}]{Yuankun}{Xie}
\small \name[affiliation={1,*}]{Ruibo}{Fu}
\small \name[affiliation={1}]{Zhengqi}{Wen}
\small \name[affiliation={4,5}]{Jianhua}{Tao}
\small \name[affiliation={1,2}]{Zhiyong}{Wang}
\small \name[affiliation={1,2}]{Xin}{Qi}
\small \name[affiliation={1}]{Xuefei}{Liu}
\small \name[affiliation={1}]{Yongwei}{Li}
\small \name[affiliation={2}]{Yukun}{Liu}
\small \name[affiliation={1,2}]{Xiaopeng}{Wang}
\small \name[affiliation={6}]{Shuchen}{Shi}

\address{
	\small $^1$Institute of Automation, Chinese Academy of Sciences 
	\small $^2$School of Artificial Intelligence, University of Chinese Academy of Sciences
	\small $^3$School of Information and Communication Engineering, Communication University of China
	\small $^4$Department of Automation, Tsinghua University
        \small $^5$ Beijing National Research Center for Information Science and Technology, Tsinghua University
        \small $^6$Shanghai Polytechnic University}
\email{luyi22@mails.ucas.ac.cn, xieyuankun@cuc.edu.cn, ruibo.fu@nlpr.ia.ac.cn}


\begin{document}

\maketitle
\renewcommand{\thefootnote}{} 
\footnotetext[1]{$\dagger$ denotes equal contribution to this work. * denotes corresponding author.}
\begin{abstract}

With the proliferation of Large Language Model (LLM) based deepfake audio, there is an urgent need for effective detection methods. Previous deepfake audio generation methods typically involve a multi-step generation process, with the final step using a vocoder to predict the waveform from hand-crafted features. However, LLM-based audio is directly generated from discrete neural codecs in an end-to-end generation process, skipping the final step of vocoder processing. This poses a significant challenge for current audio deepfake detection (ADD) models based on vocoder artifacts. To effectively detect LLM-based deepfake audio, we focus on the core of the generation process, the conversion from neural codec to waveform. We propose Codecfake dataset, which is generated by seven representative neural codec methods.  Experiment results show that codec-trained ADD models exhibit a 41.406\% reduction in average equal error rate compared to vocoder-trained ADD models on the Codecfake test set.

\end{abstract}

\section{Introduction}

With the development of Large Language Model (LLM) based audio generation models, an increasing number of deepfake audios are being produced. Traditional deepfake audio generation methods like speech synthesis typically consist of three parts: text analysis, acoustic model, and vocoder. In contrast, LLM-based audio generation models adopt only an end-to-end generation approach. They utilize the neural codec method to discretize latent representations, transforming the generation task from an autoregressive task into a discrete representation classification task, which improves efficiency and stability of the generation model. Currently, the growing prevalence of LLM-based deepfake audio poses significant challenges to current audio deepfake detection (ADD) models. 

In light of these challenge, the necessity for developing methods of detecting the LLM-based deepfake audio become increasingly apparent. Currently, researchers have undertaken a series of studies on deepfake audio detection around competitions such as ASVspoof series \cite{nautsch2021asvspoof} and ADD Challenge series \cite{yi2022add}. Advanced methods have been proposed that can achieve an intra-domain Equal Error Rate (EER) of less than 1\% \cite{eom22_interspeech, xie23c_interspeech, martin2022vicomtech}. However, these methods often experience performance degradation when faced with real-world scenarios \cite{muller22_interspeech}. This can be attributed to the inability of ADD models to generalize and detect unseen spoofing methods that were not present in the training set. 

To address the performance degradation of ADD models in the presence of unknown forgery methods, the most direct approach is data augmentation, which aims to expand the training distribution for improved generalizability. Especially in the field of ADD,
training ADD models with novel deepfake methods can significantly enhance model performance. For traditional audio deepfake generation methods, a vocoder is the core in the generation backend. Therefore, it is significant to augment the data with various vocoders to effectively detect vocoder-based deepfake audio. Wavefake \cite{frank2021wavefake} presents a vocoder-based audio deepfake dataset collected from ten sample sets across six different network architectures, spanning two languages. Wavefake datasets covers commonly used vocoders
and demonstrate that training with Wavefake dataset enables high accuracy in detecting vocoder-based deepfake audio. After Wavefake, more vocoder-based approach have been introduced to propel the development of the ADD field. Sun et al. \cite{sun2023ai} construct a vocoder-based dataset called LibriSeVoc and explore the artifact of different vocoder. More Recently, Wang et al. \cite{wang2023can} employed multiple neural vocoders to create a large-scale dataset of vocoded data and demonstrated that this approach can improve generalizability.
\begin{table}[]
	\caption{Comparision of different audio language models.}
	\fontsize{8.5}{9.8}\selectfont
	\setlength{\tabcolsep}{0.4mm}{} 
	\renewcommand{\arraystretch}{1.2}
	\label{tab:ALM}
	\begin{tabular}{|c|c|c|}
		\hline
		\textbf{ALM} & \textbf{Task}                                                                                     & \textbf{Codec} \\ \hline
		AudioLM \cite{borsos2023audiolm}      & SC, PC                                                                                            & SoundStream    \\ \hline
		AudioGen \cite{kreuk2022audiogen}     & AC                                                                                                & SoundStream    \\ \hline
		VALL-E \cite{wang2023neural}       & TTS                                                                                               & EnCodec        \\ \hline
		MusicLM \cite{agostinelli2023musiclm}      & MG                                                                                                & SoundStream    \\ \hline
		VALL-E X \cite{zhang2023speak}     & TTS, S2ST                                                                                         & EnCodec        \\ \hline
		VioLA \cite{wang2023viola}        & ASR, S2TT, TTS, MT                                                                                & EnCodec        \\ \hline
		MusicGen \cite{copet2024simple}     & MG, SG                                                                                            & EnCodec        \\ \hline
		SpeechX \cite{wang2023speechx}      & SE, SR, TSE,TTS, SPED                                                                             & EnCodec        \\ \hline
		LauraGPT \cite{chen2023lauragpt}     & \begin{tabular}[c]{@{}c@{}}ASR, S2TT, TTS, MT, \\ SE, AAC, SER, SLU\end{tabular}                  & FunCodec       \\ \hline
		UniAudio \cite{yang2023uniaudio}     & \begin{tabular}[c]{@{}c@{}}TTS, VC, SE, TSE, SVS, TTSO, \\ TTM, AUED, SD, ITTS, SPED\end{tabular} & EnCodec        \\ \hline
	\end{tabular}
\end{table}
\vspace{1mm}
\begin{figure*}[!t]
	\centering
	\includegraphics[width= 6.5in]{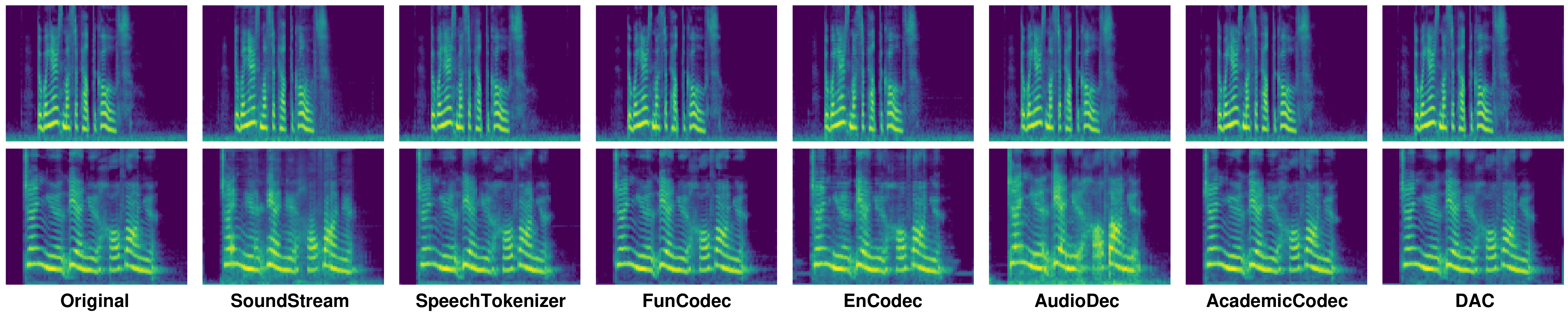}
	\hfil
	\caption{Mel-spectrogram of the original audio alongside seven codec-based audio samples generated from the original. The top row is generated from VCTK, and the bottom row is generated from AISHELL3.}
	\label{fig:mel}
\end{figure*}

The above studies have explored the core component, vocoders, of deepfake audio and introduced new datasets to better detect vocoder-based deepfake audio. However, current mainstream LLM-based deepfake audio does not utilize vocoder for waveform generation; instead, they employ neural codec approaches to generate audio from discrete codec. As illustrated in Table.~\ref{tab:ALM}, existing audio language model (ALM) universally adopt neural codecs to construct discrete encodings. Those codec-based ALMs frequently support a variety of audio generation tasks and exhibit strong generalization capabilities on unseen audio generation tasks \cite{yang2023uniaudio}.  Therefore, there is an urgent need in to establish a dataset based on neural codec for research, aiming to effectively detect LLM-based deepfake audio.

In this paper, we propose Codecfake, an initial dataset for detecting LLM-based deepfake audio, spanning two languages. We select seven representative open-source neural codec models to construct fake audio. These codec types encompass the current mainstream LLM-based audio generation models. We initially observe that the features of the original and codec-based audio exhibit only subtle differences, as shown in Figure \ref{fig:mel}. This poses a significant challenge for ADD methods. To effective detect the codec-based audio, we first investigate whether the current vocoder-trained ADD models can effectively detect codec-based audio. Then, we explore the effectiveness of codec-trained ADD models in detecting codec-based audio. Additionally, we test their performance in out-of-distribution (OOD) scenarios, specifically when faced with novel codec and detecting vocoder-based audio. Experiments demonstrate that  vocoder-trained ADD models are insufficient for effectively detecting codec-based audio. Leveraging the proposed Codecfake dataset for training, we achieve a 41.406\% reduction in EER on Codecfake test set compared to vocoder-trained ADD model.

\section{Method}
\subsection{Architectures for generating codec-based fake audio}
In this section, we provide a brief overview of the architectures and differences of the seven neural audio codecs models (F01-F07) used to construct our dataset. Those seven methods encompass all major neural audio codecs models proposed in recent years, ensuring coverage of the codecs types employed by mainstream audio language models and potentially superior neural audio codecs models that may be adopted by future audio language models.

\textbf{F01-SoundStream} \cite{zeghidour2021soundstream}. SoundStream is a milestone achievement in neural audio codecs, employing a classic architecture consisting of encoder, quantizer, and decoder components. Its quantizer utilizes residual vector quantizers (RVQ) \cite{van2017neural}, and it integrates denoising functionality throughout the entire architecture.

\textbf{F02-SpeechTokenizer} \cite{zhang2024speechtokenizer}. SpeechTokenizer utilizes semantic tokens from Hubert L9 as a teacher for the RVQ process. This aids in decoupling content information from acoustic information, with content information processed in the first layer of the tokenizer, while acoustic information is retained in subsequent layers.

\textbf{F03-FunCodec} \cite{du2023funcodec}. FunCodec encodes audio in the frequency domain, requiring fewer parameters to achieve the same effect. Additionally, it evaluates the impact of semantic information on speech codecs, thereby enhancing speech quality at low bit rates.

\textbf{F04-EnCodec} \cite{defossez2022high}. The structure of Encodec is similar to that of SoundStream. It integrates additional LSTM \cite{hochreiter1997long} layers and lightweight Transformer-based language models to model RVQ encoding, thereby accelerating inference speed while maintaining quality.

\textbf{F05-AudioDec} \cite{wu2023audiodec}. Building upon Encodec, AudioDec introduces improvements by employing a group convolution mechanism to accelerate inference speed for real-time transmission. Additionally, it incorporates a multi-period discriminator from HiFi-GAN \cite{su2020hifi} to ensure the generation of high-fidelity audio.

\textbf{F06-AcademicCodec} \cite{yang2023hifi}. AcademiCodec introduces group-residual vector quantization and utilizes multiple parallel RVQ groups, aiming to enhance audio reconstruction performance within the constraints of codebook size.

\textbf{F07-Descript-audio-codec (DAC)} \cite{kumar2024high}. DAC is a versatile neural codec model capable of maintaining high-quality audio across a wide frequency range. It employs enhanced residual vector quantization and utilizes periodic activation functions 	\cite{ziyin2020neural}.

\begin{table*}[t]
	\caption{Number of samples in each subset of Codecfake dataset.}
	\label{tab:number}
	\centering
	\setlength{\tabcolsep}{2mm} 
	\renewcommand{\arraystretch}{1.2}
	\begin{tabular}{|c|c|c|c|c|c|c|c|c|c|}
		\hline
		\multirow{2}{*}{Type}&\multicolumn{8}{c|}{Seen}&Unseen \\
		\cline{2-10}

		 &Train &Dev &C1&C2&C3&C4 &C5&C6&C7\\
		\hline
		Real  &105821 &13228 &13228&13228 &13228&13228&13228&13228&13228\\
		\hline
		Fake  &634926 &79368 &13228&13228 &13228&13228&13228&13228&132277\\
		\hline
		Total  &740747 &92596 &26456 &26456 &26456&26456 &26456&26456&145505\\
		\hline
	\end{tabular}
	
\end{table*}

\begin{table}[t]
	\caption{Neural codec models detail.}
	\label{tab:encodec}
	\renewcommand{\arraystretch}{1.0}
	\begin{tabular}{|c|c|c|c|c|}
		\hline
		\textbf{Type} &\textbf{Codec}& \textbf{SR} & \textbf{BPS} & \textbf{Quantizers} \\ \hline
		F01&SoundStream    & 16k & 4k & 8                                            \\ \hline
		F02&SpeechToknizer & 16k & 4k & 8                                           \\ \hline
		F03&FuncCodec      & 16k & 16k & 32                                        \\ \hline
		F04&EnCodec        & 24k & 6k & 8                                           \\ \hline
		F05&AudioDec       & 24k & 6.4k & 8                                        \\ \hline
		F06&AcademicCodec  & 24k & 3k & 4                                           \\ \hline
		F07&DAC            & 44k & 8k & 9                                        \\ \hline
		
	\end{tabular}
\end{table}

\subsection{The Generation Process of codec-based fake audio}
In this section, we will introduce the training details of neural audio codecs, the datasets used, as well as the inference process and the construction of the Codecfake dataset.

During the training phase, neural codec models are trained using the LibriTTS \cite{zen2019libritts} dataset, which is a multi-speaker English corpus comprising approximately 585 hours of read English speech at a sampling rate of 24kHz. The LibriTTS corpus is specifically designed for research in text-to-speech systems and is derived from the original materials of the LibriSpeech \cite{panayotov2015librispeech} corpus. We typically set the parameters of each model to the default values mentioned in the original paper or to parameter settings widely adopted by audio language models for training, continuing training until the model converges.

During the inference phase, we utilize the trained seven neural codec models to re-encode and decode data from the VCTK \cite{Veaux2017CSTRVC} (This CSTR VCTK Corpus includes speech data uttered by 110 English speakers with various accents.) and AISHELL3 \cite{shi21c_interspeech} (The corpus contains roughly 85 hours of emotion-neutral recordings spoken by 218 native Chinese mandarin speakers) datasets separately. More neural codec models details is ilustrated in Table \ref{tab:encodec}.

\section{Experiments}
\subsection{Data details}
Our proposed Codecfake dataset consists of 1,058,216 audio samples, including 132,277 real samples (44,242 samples from VCTK and 88,035 samples from AISHELL3) and 925,939 fake samples generated by seven different codec methods. Regarding ADD experiments, we divided the 132,277 source real samples into a training subset of 105,821 samples, a development subset of 13,228 samples, and an evaluation subset of 13,228 samples. The same dataset splitting strategy was applied to the fake audio generated using the six different forgery methods (F01-F06). We designated F07 as an unseen fake methods for the generalization verification of the ADD model.  This implies that there are no F07 utterances in either the training or development sets. The training set of Codecfake comprises a total of 740,747 samples, and the development set contains 92,596 samples. For evaluation, we established seven conditions (C1-C7) for codec methods F01-F07, with each condition including the evaluation subset of real audio. Details are provided in Table \ref{tab:number}.

\subsection{Implementation details}
We comprehensively evaluated the Codecfake dataset using state-of-the-art ADD methods, namely AASIST \cite{jung2022aasist} and LCNN \cite{lavrentyeva19_interspeech}. In terms of features, we utilized the hand-crafted feature mel-spectrogram and self-supervised feature wav2vec2 (W2V2) representations. For the mel-spectrogram, we extracted a 80-dimensional mel-spectrogram. For W2V2, we employed the Wav2Vec-XLS-R\footnote{https://huggingface.co/facebook/wav2vec2-xls-r-300m} model with frozen parameters, extracting the 1024-dimensional hidden states as the feature representation. All audio samples were downsampled to 16000 Hz and trimmed or padded to a duration of 4 seconds.
All models used Adam optimizer with a learning rate of $10^{-4}$ and cosine annealing learning rate decay. We conducted training for 100 epochs using weighted cross-entropy, with the weight for the real class set to 10 and for the fake class set to 1. For the codec-trained ADD model, training was performed for 10 epochs. The model that exhibited the best performance on the validation set was selected as the evaluation model.

For experimental evaluation, we utilized the official implementation for EER calculation\footnote{https://github.com/asvspoof-challenge/2021/blob/main/eval-package/eval\_metrics.py}, maintaining precision up to three decimal places.
For the computation of the confusion matrix, we utilized a threshold of 0.5 for discriminating between real and fake predictions.

\begin{table*}[t]
	\caption{EER (\%) results for ADD model trained by 19LA training set. AVG represents the average EER across C1-C7.}
	\label{tab:vocoder-trained}
	\centering
	\setlength{\tabcolsep}{2mm} 
	\renewcommand{\arraystretch}{1.2}
	\begin{tabular}{|c|c|c|c|c|c|c|c|c|c|}
		\hline
		Model &19LA  &C1&C2&C3&C4 &C5 &C6 &C7 &AVG\\
		\hline
		Mel-LCNN  &5.084		&36.127	&49.304	&\bf 39.234	&49.160	&46.991	&\bf 36.967	&49.584	 &43.910 \\
		\hline
		W2V2-LCNN  &0.625		&\bf 19.481	&45.260	&45.010	&\bf 20.683	&\bf 34.321	&44.390	&\bf 43.717	 &\bf 36.123
		\\
		\hline
		W2V2-AASIST  &\bf 0.122  &40.142	&\bf 42.908	&44.564	&33.580	&39.197	&44.889	&45.804	&41.583  \\
		\hline
		
	\end{tabular}
	
\end{table*}

\begin{table*}[t]
	\caption{EER (\%) results for ADD model trained by Codecfake training set. AVG represents the average EER across C1-C7.}
	\label{tab:codec-trained}
	\centering
	\setlength{\tabcolsep}{2.5mm} 
	\renewcommand{\arraystretch}{1.2}
	\begin{tabular}{|c|c|c|c|c|c|c|c|c|c|}
		\hline
		Model &19LA  &C1&C2&C3&C4 &C5 &C6 &C7  &AVG\\
		\hline
		Mel-LCNN  &26.826	&2.147	&9.563	&8.618	&12.239	&20.033	&8.951	&30.867	 &  13.203
		\\
		\hline
		W2V2-LCNN  &4.433		&\bf 0.030	&0.151	&0.144	&0.113	&0.635	&2.185	&5.178	 &1.205
		\\
		\hline
		W2V2-AASIST  &\bf 3.806 &0.167	&\bf 0.008	&\bf 0.023	&\bf 0.015	&\bf 0.038	&\bf 0.106	&\bf 0.884 &\bf 0.177
		\\
		\hline
	\end{tabular}
	
\end{table*}
\begin{figure*}[!t]
	\centering
	\includegraphics[width= 5.5in]{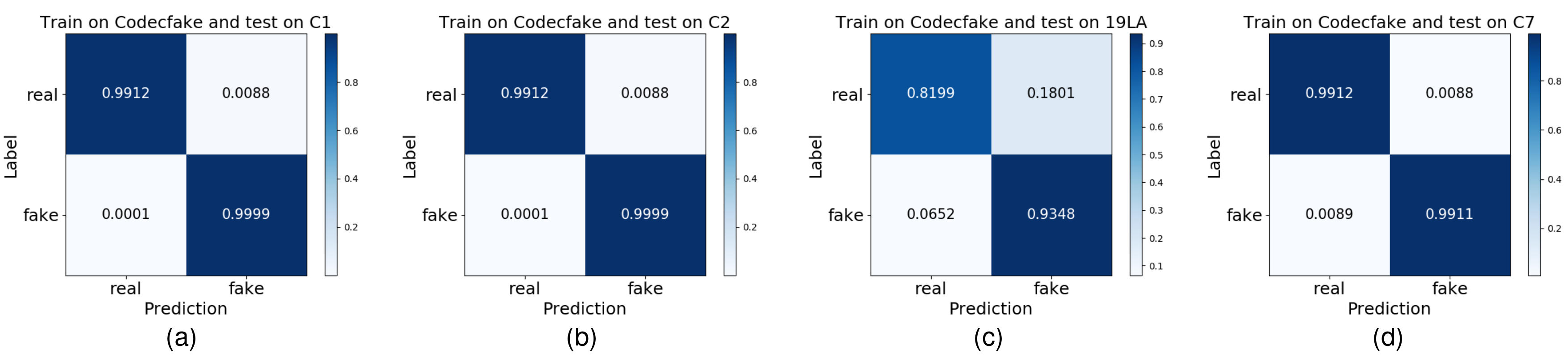}
	\hfil
	\caption{The confusion matrices under different test conditions. (a), (b), (c), and (d) correspond to W2V2-AASIST trained on the Codecfake training set and tested on C3, C6, 19LA, and C7, respectively. }
	\label{fig:confusion}
\end{figure*}

\section{Results and Discussion}
\subsection{Vocoder-trained ADD models results}
First, we explore whether the vocoder-trained ADD model can effectively detect codec-based audio, as outlined in Table \ref{tab:vocoder-trained}. Concretely, we conducted training on three distinct models: Mel-LCNN, W2V2-LCNN and W2V2-AASIST, utilizing the ASVspoof2019LA (19LA) \cite{todisco19_interspeech} training set. From the results, we can first observe that they achieve promising result in 19LA test sets and W2V2-AASIST achieve the lowest EER with 0.122\%. As the 19LA training set incorporates six spoofing methods, and the test set encompasses a total of 19 methods, achieving favorable results on the 19LA test set suggests that the detection system exhibits generalizability. However, when the vocoder-trained ADD models evaluate on the Codecfake test sets, the results are quite unfavorable. Mel-LCNN, W2V2-LCNN, and W2V2-AASIST achieve average EER of 43.910\%, 36.123\%, and 41.583\%, respectively, which represents a decrease in performance compared to the 19LA test set, with increases of 38.826\%, 35.498\%, and 41.461\%, respectively. This indicates that models trained with vocoders cannot generalize to effectively detect codec-based audio. Furthermore, it also suggests that codec-based audio exhibits minimal differences from real speech, posing a significant challenge to the field of ADD.

\subsection{Codec-trained ADD models results}
Due to the poor performance of vocoder-trained ADD models on the Codecfake test set, we proceed to train the baseline models using the Codecfake training set. The results are presented in the Table \ref{tab:codec-trained}. Mel-LCNN, W2V2-LCNN, and W2V2-AASIST achieve average EER of 13.203\%, 1.205\%, and 0.177\%, exhibited reductions in average EER of 30.707\%, .34.918\%, and 41.406\%, respectively, compared to the vocoder-trained model. 
All ADD models trained with the Codecfake dataset exhibit promising results under seen conditions C1-C6. Among these, W2V2-AASIST achieved the best average EER of 0.177\%. In the unseen codec test condition C7, W2V2-AASIST achieves an EER of 0.884\%, indicating that training with the Codecfake dataset can generalize to detecting unknown codec-based audio. Furthermore, we found that using only the codec-trained ADD model can detect vocoder-based audio. Specifically, W2V2-AASIST achieves a 3.806\% EER on the 19LA test set.

Through the experiments, we observed a decrease in performance of vocoder-trained ADD models when testing codec-based audio, and similarly, a decrease in performance of codec-trained ADD models when testing vocoder-based audio such as test on 19LA. To delve deeper into the factors contributing to the degraded performance of ADD models under OOD test conditions, we plotted the confusion matrix graphs in Figure \ref*{fig:confusion}. We first plotted the confusion matrices for the best results in ID condition as shown in Figure \ref*{fig:confusion}(a) and Figure \ref*{fig:confusion}(b). It can be observed that both real and fake samples are nearly accurately predicted. Particularly for fake speech, 99.99\% of the fake audio in both C1 and C2 is precisely classified as fake. Then, in Figure \ref*{fig:confusion}(c) and Figure \ref*{fig:confusion}(d), the codec-trained ADD model demonstrates an enhanced ability to distinguish between real and fake audio. This is particularly evident with codec-based audio, which is inherently more authentic than vocoder-based audio, leading to superior performance in both the OOD scenario of 19LA and the unseen condition of C7.

\subsection{Discussion}
Through our experiments, we have identified some limitations and areas for further improvement in our work. Future efforts will focus on the following items to make enhancements.
\begin{itemize}
	\item {{\bf More diverse LLM-based audio Dataset}: Codecfake incorporates mainstream codec methods. However, training on Codecfake still performs inadequately on unknown codec methods. Therefore, there is a need to establish a diverse codec dataset to encompass nerual codec generation methods in ALM. }
	\item {{\bf Influential factors of codec-based audio}: The generation of codec-based audio is influenced by various parameters, including the number of quantizers, bps, and other codec-specific settings. Exploring the impact of these parameters is crucial for effectively detecting codec-based audio.}
	\item {{\bf Generalized ADD methods}: The current ADD methods, such as W2V2-AASIST, exhibit a significant performance decline in OOD scenarios. Designing a generalized approach that can detect both vocoder-based audio and LLM-based audio is currently a crucial challenge.}	
	\item {{\bf Source tracing methods}: It is crucial to develop source tracing methods, as this significantly contributes to safeguarding the copyright of ALM developers. Specifically, we need to design a source tracing method with generalization capabilities. This method should be able to perform classification in the ID ALM and identify novel ALM in OOD situations, which can effectively addressing the continuous emergence of ALM algorithms.}

\end{itemize}
\section{Conclusion}
In this paper, we propose Codecfake dataset for detect the audio based on LLM. We select seven representative neural codec methods to construt fake audio. With this dataset, we first investigate the performance of vocoder-trained ADD models for codec deepfake detection task. Due to the unsatisfactory outcomes achieved by vocoder-trained ADD methods, we turned to training with the Codecfake dataset. The experimental results underscore the efficacy of training with the Codecfake dataset, as the W2V2-AASIST model attains an average lowest EER of 0.177\%, reflecting a significant decrease of 41.406\% compared to the vocoder-trained models. 

\section{Acknowledgements}
This work is supported by the National Natural Science Foundation of China (NSFC) (No.62101553, No.62306316, No.U21B20210, No. 62201571).

%
%

\tiny
\bibliographystyle{IEEEtran}
\bibliography{mybib}

\begin{thebibliography}{10}
\providecommand{\url}[1]{#1}
\csname url@samestyle\endcsname
\providecommand{\newblock}{\relax}
\providecommand{\bibinfo}[2]{#2}
\providecommand{\BIBentrySTDinterwordspacing}{\spaceskip=0pt\relax}
\providecommand{\BIBentryALTinterwordstretchfactor}{4}
\providecommand{\BIBentryALTinterwordspacing}{\spaceskip=\fontdimen2\font plus
\BIBentryALTinterwordstretchfactor\fontdimen3\font minus
  \fontdimen4\font\relax}
\providecommand{\BIBforeignlanguage}[2]{{%
\expandafter\ifx\csname l@#1\endcsname\relax
\typeout{** WARNING: IEEEtran.bst: No hyphenation pattern has been}%
\typeout{** loaded for the language `#1'. Using the pattern for}%
\typeout{** the default language instead.}%
\else
\language=\csname l@#1\endcsname
\fi
#2}}
\providecommand{\BIBdecl}{\relax}
\BIBdecl

\bibitem{nautsch2021asvspoof}
A.~Nautsch, X.~Wang, N.~Evans, T.~H. Kinnunen, V.~Vestman, M.~Todisco,
  H.~Delgado, M.~Sahidullah, J.~Yamagishi, and K.~A. Lee, ``Asvspoof 2019:
  spoofing countermeasures for the detection of synthesized, converted and
  replayed speech,'' \emph{IEEE Transactions on Biometrics, Behavior, and
  Identity Science}, vol.~3, no.~2, pp. 252--265, 2021.

\bibitem{liu2023asvspoof}
X.~Liu, X.~Wang, M.~Sahidullah, J.~Patino, H.~Delgado, T.~Kinnunen, M.~Todisco,
  J.~Yamagishi, N.~Evans, A.~Nautsch \emph{et~al.}, ``Asvspoof 2021: Towards
  spoofed and deepfake speech detection in the wild,'' \emph{IEEE/ACM
  Transactions on Audio, Speech, and Language Processing}, 2023.

\bibitem{yi2022add}
J.~Yi, R.~Fu, J.~Tao, S.~Nie, H.~Ma, C.~Wang, T.~Wang, Z.~Tian, Y.~Bai, C.~Fan
  \emph{et~al.}, ``Add 2022: the first audio deep synthesis detection
  challenge,'' in \emph{Proceedings of ICASSP}.\hskip 1em plus 0.5em minus
  0.4em\relax IEEE, 2022, pp. 9216--9220.

\bibitem{eom22_interspeech}
Y.~Eom, Y.~Lee, J.~S. Um, and H.~R. Kim, ``{Anti-Spoofing Using Transfer
  Learning with Variational Information Bottleneck},'' in \emph{Proc.
  Interspeech 2022}, 2022, pp. 3568--3572.

\bibitem{martin2022vicomtech}
J.~M. Mart{\'\i}n-Do{\~n}as and A.~{\'A}lvarez, ``The vicomtech audio deepfake
  detection system based on wav2vec2 for the 2022 add challenge,'' in
  \emph{ICASSP 2022-2022 IEEE International Conference on Acoustics, Speech and
  Signal Processing (ICASSP)}.\hskip 1em plus 0.5em minus 0.4em\relax IEEE,
  2022, pp. 9241--9245.

\bibitem{muller22_interspeech}
N.~Müller, P.~Czempin, F.~Diekmann, A.~Froghyar, and K.~Böttinger, ``{Does
  Audio Deepfake Detection Generalize?}'' in \emph{Proc. Interspeech 2022},
  2022, pp. 2783--2787.

\bibitem{frank2021wavefake}
J.~Frank and L.~Sch{\"o}nherr, ``{WaveFake: A Data Set to Facilitate Audio
  Deepfake Detection},'' in \emph{Thirty-fifth Conference on Neural Information
  Processing Systems Datasets and Benchmarks Track}, 2021.

\bibitem{sun2023ai}
C.~Sun, S.~Jia, S.~Hou, and S.~Lyu, ``Ai-synthesized voice detection using
  neural vocoder artifacts,'' in \emph{Proceedings of the IEEE/CVF Conference
  on Computer Vision and Pattern Recognition}, 2023, pp. 904--912.

\bibitem{wang2023can}
X.~Wang and J.~Yamagishi, ``Can large-scale vocoded spoofed data improve speech
  spoofing countermeasure with a self-supervised front end?'' \emph{arXiv
  preprint arXiv:2309.06014}, 2023.

\bibitem{borsos2023audiolm}
Z.~Borsos, R.~Marinier, D.~Vincent, E.~Kharitonov, O.~Pietquin, M.~Sharifi,
  D.~Roblek, O.~Teboul, D.~Grangier, M.~Tagliasacchi \emph{et~al.}, ``Audiolm:
  a language modeling approach to audio generation,'' \emph{IEEE/ACM
  Transactions on Audio, Speech, and Language Processing}, 2023.

\bibitem{kreuk2023audiogen}
\BIBentryALTinterwordspacing
F.~Kreuk, G.~Synnaeve, A.~Polyak, U.~Singer, A.~D{\'e}fossez, J.~Copet,
  D.~Parikh, Y.~Taigman, and Y.~Adi, ``Audiogen: Textually guided audio
  generation,'' in \emph{The Eleventh International Conference on Learning
  Representations}, 2023. [Online]. Available:
  \url{https://openreview.net/forum?id=CYK7RfcOzQ4}
\BIBentrySTDinterwordspacing

\bibitem{wang2023neural}
C.~Wang, S.~Chen, Y.~Wu, Z.~Zhang, L.~Zhou, S.~Liu, Z.~Chen, Y.~Liu, H.~Wang,
  J.~Li \emph{et~al.}, ``Neural codec language models are zero-shot text to
  speech synthesizers,'' \emph{arXiv preprint arXiv:2301.02111}, 2023.

\bibitem{agostinelli2023musiclm}
A.~Agostinelli, T.~I. Denk, Z.~Borsos, J.~Engel, M.~Verzetti, A.~Caillon,
  Q.~Huang, A.~Jansen, A.~Roberts, M.~Tagliasacchi \emph{et~al.}, ``Musiclm:
  Generating music from text,'' \emph{arXiv preprint arXiv:2301.11325}, 2023.

\bibitem{zhang2023speak}
Z.~Zhang, L.~Zhou, C.~Wang, S.~Chen, Y.~Wu, S.~Liu, Z.~Chen, Y.~Liu, H.~Wang,
  J.~Li \emph{et~al.}, ``Speak foreign languages with your own voice:
  Cross-lingual neural codec language modeling,'' \emph{arXiv preprint
  arXiv:2303.03926}, 2023.

\bibitem{wang2023viola}
T.~Wang, L.~Zhou, Z.~Zhang, Y.~Wu, S.~Liu, Y.~Gaur, Z.~Chen, J.~Li, and F.~Wei,
  ``Viola: Unified codec language models for speech recognition, synthesis, and
  translation,'' \emph{arXiv preprint arXiv:2305.16107}, 2023.

\bibitem{copet2024simple}
J.~Copet, F.~Kreuk, I.~Gat, T.~Remez, D.~Kant, G.~Synnaeve, Y.~Adi, and
  A.~D{\'e}fossez, ``Simple and controllable music generation,'' \emph{Advances
  in Neural Information Processing Systems}, vol.~36, 2024.

\bibitem{wang2023speechx}
X.~Wang, M.~Thakker, Z.~Chen, N.~Kanda, S.~E. Eskimez, S.~Chen, M.~Tang,
  S.~Liu, J.~Li, and T.~Yoshioka, ``Speechx: Neural codec language model as a
  versatile speech transformer,'' \emph{arXiv preprint arXiv:2308.06873}, 2023.

\bibitem{chen2023lauragpt}
Q.~Chen, Y.~Chu, Z.~Gao, Z.~Li, K.~Hu, X.~Zhou, J.~Xu, Z.~Ma, W.~Wang, S.~Zheng
  \emph{et~al.}, ``Lauragpt: Listen, attend, understand, and regenerate audio
  with gpt,'' \emph{arXiv preprint arXiv:2310.04673}, 2023.

\bibitem{yang2023uniaudio}
D.~Yang, J.~Tian, X.~Tan, R.~Huang, S.~Liu, X.~Chang, J.~Shi, S.~Zhao, J.~Bian,
  X.~Wu \emph{et~al.}, ``Uniaudio: An audio foundation model toward universal
  audio generation,'' \emph{arXiv preprint arXiv:2310.00704}, 2023.

\bibitem{zen2019libritts}
H.~Zen, V.~Dang, R.~Clark, Y.~Zhang, R.~J. Weiss, Y.~Jia, Z.~Chen, and Y.~Wu,
  ``Libritts: A corpus derived from librispeech for text-to-speech,'' in
  \emph{Proc. Interspeech}, Sep. 2019.

\bibitem{panayotov2015librispeech}
V.~Panayotov, G.~Chen, D.~Povey, and S.~Khudanpur, ``Librispeech: an asr corpus
  based on public domain audio books,'' in \emph{2015 IEEE international
  conference on acoustics, speech and signal processing (ICASSP)}.\hskip 1em
  plus 0.5em minus 0.4em\relax IEEE, 2015, pp. 5206--5210.

\bibitem{Veaux2017CSTRVC}
C.~Veaux, J.~Yamagishi, and K.~MacDonald, ``Cstr vctk corpus: English
  multi-speaker corpus for cstr voice cloning toolkit,'' 2017.

\bibitem{shi21c_interspeech}
Y.~Shi, H.~Bu, X.~Xu, S.~Zhang, and M.~Li, ``{AISHELL-3: A Multi-Speaker
  Mandarin TTS Corpus},'' in \emph{Proc. Interspeech 2021}, 2021, pp.
  2756--2760.

\bibitem{zeghidour2021soundstream}
N.~Zeghidour, A.~Luebs, A.~Omran, J.~Skoglund, and M.~Tagliasacchi,
  ``Soundstream: An end-to-end neural audio codec,'' \emph{IEEE/ACM
  Transactions on Audio, Speech, and Language Processing}, vol.~30, pp.
  495--507, 2021.

\bibitem{van2017neural}
A.~Van Den~Oord, O.~Vinyals \emph{et~al.}, ``Neural discrete representation
  learning,'' \emph{Advances in neural information processing systems},
  vol.~30, 2017.

\bibitem{zhang2024speechtokenizer}
\BIBentryALTinterwordspacing
X.~Zhang, D.~Zhang, S.~Li, Y.~Zhou, and X.~Qiu, ``Speechtokenizer: Unified
  speech tokenizer for speech language models,'' in \emph{The Twelfth
  International Conference on Learning Representations}, 2024. [Online].
  Available: \url{https://openreview.net/forum?id=AF9Q8Vip84}
\BIBentrySTDinterwordspacing

\bibitem{du2023funcodec}
Z.~Du, S.~Zhang, K.~Hu, and S.~Zheng, ``Funcodec: A fundamental, reproducible
  and integrable open-source toolkit for neural speech codec,'' \emph{arXiv
  preprint arXiv:2309.07405}, 2023.

\bibitem{defossez2022high}
A.~D{\'e}fossez, J.~Copet, G.~Synnaeve, and Y.~Adi, ``High fidelity neural
  audio compression,'' \emph{arXiv preprint arXiv:2210.13438}, 2022.

\bibitem{hochreiter1997long}
S.~Hochreiter and J.~Schmidhuber, ``Long short-term memory,'' \emph{Neural
  computation}, vol.~9, no.~8, pp. 1735--1780, 1997.

\bibitem{wu2023audiodec}
Y.-C. Wu, I.~D. Gebru, D.~Markovi{\'c}, and A.~Richard, ``Audiodec: An
  open-source streaming high-fidelity neural audio codec,'' in \emph{ICASSP
  2023-2023 IEEE International Conference on Acoustics, Speech and Signal
  Processing (ICASSP)}.\hskip 1em plus 0.5em minus 0.4em\relax IEEE, 2023, pp.
  1--5.

\bibitem{su2020hifi}
J.~Su, Z.~Jin, and A.~Finkelstein, ``Hifi-gan: High-fidelity denoising and
  dereverberation based on speech deep features in adversarial networks,''
  \emph{arXiv preprint arXiv:2006.05694}, 2020.

\bibitem{yang2023hifi}
D.~Yang, S.~Liu, R.~Huang, J.~Tian, C.~Weng, and Y.~Zou, ``Hifi-codec:
  Group-residual vector quantization for high fidelity audio codec,''
  \emph{arXiv preprint arXiv:2305.02765}, 2023.

\bibitem{kumar2024high}
R.~Kumar, P.~Seetharaman, A.~Luebs, I.~Kumar, and K.~Kumar, ``High-fidelity
  audio compression with improved rvqgan,'' \emph{Advances in Neural
  Information Processing Systems}, vol.~36, 2024.

\bibitem{ziyin2020neural}
L.~Ziyin, T.~Hartwig, and M.~Ueda, ``Neural networks fail to learn periodic
  functions and how to fix it,'' \emph{Advances in Neural Information
  Processing Systems}, vol.~33, pp. 1583--1594, 2020.

\bibitem{jung2022aasist}
J.-w. Jung, H.-S. Heo, H.~Tak, H.-j. Shim, J.~S. Chung, B.-J. Lee, H.-J. Yu,
  and N.~Evans, ``Aasist: Audio anti-spoofing using integrated spectro-temporal
  graph attention networks,'' in \emph{Proceedings of the ICASSP}, 2022, pp.
  6367--6371.

\bibitem{lavrentyeva19_interspeech}
G.~Lavrentyeva, S.~Novoselov, A.~Tseren, M.~Volkova, A.~Gorlanov, and
  A.~Kozlov, ``{STC Antispoofing Systems for the ASVspoof2019 Challenge},'' in
  \emph{Proc. Interspeech 2019}, 2019, pp. 1033--1037.

\bibitem{todisco19_interspeech}
M.~Todisco, X.~Wang, V.~Vestman, M.~Sahidullah, H.~Delgado, A.~Nautsch,
  J.~Yamagishi, N.~Evans, T.~H. Kinnunen, and K.~A. Lee, ``{ASVspoof 2019:
  Future Horizons in Spoofed and Fake Audio Detection},'' in \emph{Proc.
  Interspeech 2019}, 2019, pp. 1008--1012.

\end{thebibliography}

\end{document}